\documentclass[12pt,journal,draftclsnofoot,onecolumn]{IEEEtran}

\usepackage{mathptmx}   
\usepackage{amsfonts, amssymb, amsmath}
\usepackage{epsfig}
\usepackage{theorem}
\usepackage{textcomp}
\usepackage{tipa}
\usepackage{multirow}

\makeatletter
\let\NAT@parse\undefined
\makeatletter
\usepackage{natbib}

\newtheorem{theorem}{Theorem}
\newtheorem{proposition}[theorem]{Proposition}
\newtheorem{corollary}[theorem]{Corollary}
\newtheorem{lemma}[theorem]{Lemma}
\newtheorem{problem}{Problem}
\newcommand{\qed}{\hfill $\Box$\\}

\def\pmt{PMT}
\def\pmg{PMG}
\def\pmr{PMR}

\def\G{\mathbf G}

\newcommand*{\FULLPAPER}{}%

\begin{document}
\title{Pebble Motion on Graphs with Rotations: Efficient Feasibility Tests and Planning Algorithms}
\author{
\begin{tabular}{c}
Jingjin~Yu,~Daniela~Rus
\end{tabular}
\thanks{Jingjin Yu and Daniela Rus are the Computer Science and Artificial Intelligence Lab at the Massachusetts Institute of Technology. E-mail: \{jingjin, rus\}@csail.mit.edu. } 
}
\maketitle

\begin{abstract}We study the problem of planning paths for $p$ distinguishable pebbles (robots) residing on the vertices of an $n$-vertex connected graph with $p \le n$. A pebble may move from a vertex to an adjacent one in a time step provided that it does not collide with other pebbles. When $p = n$, the only collision free moves are synchronous rotations of pebbles on disjoint cycles of the graph. We show that the feasibility of such problems is intrinsically determined by the diameter of a (unique) permutation group induced by the underlying graph. Roughly speaking, the diameter of a group $\G$ is the minimum length of the generator product required to reach an arbitrary element of $\G$ from the identity element. Through bounding the diameter of this associated permutation group, which assumes a maximum value of $O(n^2)$, we establish a linear time algorithm for deciding the feasibility of such problems and an $O(n^3)$ algorithm for planning complete paths. 
\end{abstract}

\section{Introduction}\label{sec:intro}
\begin{figure}[!ht]
\vspace*{-5mm}
\begin{center}
  \begin{tabular}{ccc}
    \includegraphics[width=0.15\textwidth]{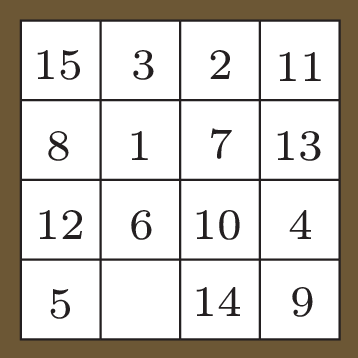} & \hspace{20mm} &
    \includegraphics[width=0.15\textwidth]{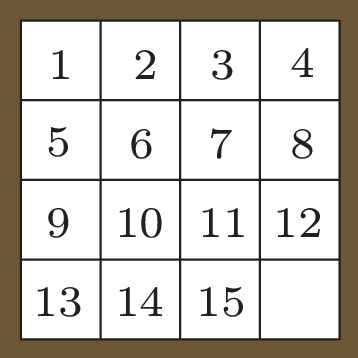}  \\
    (a) & & (b)\\
  \end{tabular}
\end{center}
\vspace*{-6mm}
\caption{\label{fig:15-puzzle} Two 15-puzzle instances. a) An unsolved instance. In the next step, one of the blocks 5, 6, 14 may move to the vacant cell, leaving behind it another vacant cell for the next move. b) The solved instance.}
\end{figure}

\vspace*{-2mm}
In Sam Loyd's 15-puzzle \cite{Loy59}, a player arranges square blocks labeled 1-15, scrambled on a $4 \times 4$ board, to achieve a shuffled row major ordering of the blocks using one empty swap cell (see, {\em e.g.}, Fig. \ref{fig:15-puzzle}). Generalizing the grid-based board to an arbitrary connected graph over $n$ vertices, the 15-puzzle becomes the problem of {\em pebble motion on graphs} (\pmg). Here, up to $n-1$ uniquely labeled pebbles on the vertices of the graph must be moved to some desired goal configuration, using unoccupied (empty) vertices as swap spaces.\footnote{We use {\em pebble} in place of {\em robot} in this paper to keep the notations consistent with \cite{AulMonParPer99,KorMilSpi84}, on which the current paper is partially based.} Since the initial work by Kornhauser et al. \cite{KorMilSpi84}, \pmg\, and its optimal variants has received significant attention in robotics \cite{SolHal12,BerSnoLinMan09,WagChoC11} and artificial intelligence \cite{KroLunBek13,StaKor11}, among others. The connection between \pmg\, and multi-robot path planning is immediately clear, with potential applications towards micro-fluidics \cite{GriAke05}, multi-robot path planning \cite{SolHal12}, and modular robot reconfiguration \cite{ReiSle06}, to name a few. 

As early as 1879, Story \cite{Sto1879} observed that the parity of a 15-puzzle instance decides whether it is solvable. Wilson \cite{Wil74} formalized this observation by showing that the reachable configurations of a 15-puzzle form an alternating group on 15 letters. An associated planning algorithm was also provided. Kornhauser et al. \cite{KorMilSpi84} improved the potentially exponential time algorithm from \cite{Wil74} by giving an algorithm for \pmg\, that runs in $O(n^3)$ time for graphs with $n$ vertices and up to $n-1$ pebbles. Auletta et al. \cite{AulMonParPer99} showed that for trees, deciding whether an instance of the pebble motion problem is feasible can be done in linear time. Recently, the linear feasibility result was extended to general graphs for \pmg\, \cite{GorHas10,YuArxiv-1301-2342}. Although not a focus of this paper, we note that computing optimal plans for such problems is generally NP-complete \cite{Gol84,RatWar90,Sur10,YuLav13AAAI}.
\vspace*{-5mm}

As evident from the techniques used in \cite{KorMilSpi84,Wil74}, \pmg\, and related problems are closely related to structures of {\em permutation groups}. Fixing a graph and the number of pebbles, and viewing the pebble moving operations as {\em generators}, all configurations reachable from an initial configuration form a group that is isomorphic to a subgroup of $\mathbf{S_n}$, the symmetric group on $n$ letters. Deciding whether a problem instance is feasible is then equivalent to deciding whether the final configuration is reachable from the initial configuration via generator products. Another interesting problem in this domain is the study of the {\em diameter} of such groups, which is the length of the longest minimal generator product required to reach a group element. Driscoll and Furst \cite{DriFur83,DriFur87} showed that any group represented by generators that are cycles of bounded degree has a diameter of $O(n^2)$ and such a generator sequence is efficiently computable. For generators of unbounded size, Babai et al. \cite{BabBeaSer04} proved that if one of the generators fixes at least $67\%$ of the domain, then the resulting group has a polynomial diameter. In contrast, groups with super polynomial diameters exist \cite{DriFur83}. 
\begin{figure}[!ht]
\vspace*{-5mm}
\begin{center}
\begin{tabular}{ccc}
    \includegraphics[width=0.25\textwidth]{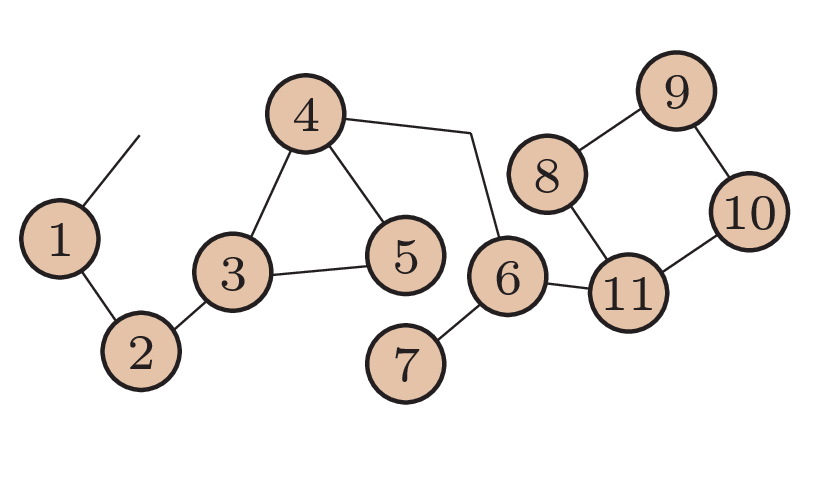}  & \hspace{20mm} &
    \includegraphics[width=0.25\textwidth]{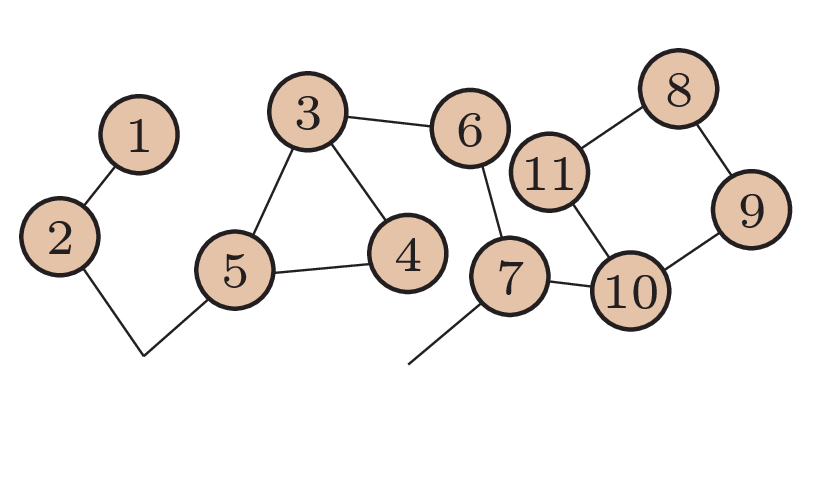} \\
    (a) & & (b)
\end{tabular}
\end{center}
\vspace*{-6mm}
\caption{\label{fig:synchro} Two configurations that can be turned into each other in a single synchronized move.}
\end{figure}
\vspace*{-5mm}

Somewhat surprisingly, a natural generalization of \pmg\, allowing rotations of the pebbles without empty swap vertices has not received much attention, possibly due to its difficulty. As an example, in Fig.~\ref{fig:synchro}(a), the pebbles labeled $3, 4$, and $5$ are allowed to rotate clockwise along the (only) triangle to achieve the configuration in Fig.~\ref{fig:synchro}(b). We call this generalization the problem of {\em pebble motion with rotations} (\pmr), a formal definition of which will follow shortly. Synchronous rotations are important to have in a multi-robot setting for at least two reasons. First, with communication, robots are able to execute synchronous rotational moves easily. Disabling such moves thus wastes robots' capabilities. Second, allowing rotational moves could allow more problem instances to be solved and could also significantly reduce the length of plans (note that the length of a plan can never be increased by adding more modes of motion). 

In this paper, we employ a group theoretic approach to derive a linear time algorithm for testing the feasibility of a given \pmr\, instance. The algorithm also implies a cubic time algorithm for computing full plans when a \pmr\, instance is feasible. Thus, we establish that \pmr\, induces similar algorithmic complexity as \pmg\, does in the sense that planning and feasibility test take $O(n^3)$ and linear time, respectively. Nevertheless, the algorithms for solving \pmg\, and \pmr\, have significant differences due to the introduction of synchronous pebble rotations. By delivering these algorithms for \pmr, we also bring forth the contribution of providing a now fairly complete landscape over graph-based multi-robot path planning problems. 

We formally define \pmg\, and \pmr\, problems in Section~\ref{sec:formulation}. In Section \ref{sec:upper}, we look at the groups generated by cyclic rotations of labeled pebbles, on graphs fully occupied by pebbles. We show that such groups have $O(n^2)$ diameters. With this intermediate result, we continue to show, in Section \ref{sec:linear}, that the feasibility test of the \pmr\, problem can be performed in $O(|V| + |E|)$ time, which implies an $O(n^3)$ algorithm for computing a feasible solution (the set of movements). We conclude the paper in Section \ref{sec:conclusion}.\footnote{Given the limited space, we focus on establishing the theoretical foundations behind the algorithms instead of the algorithms themselves. We believe such coverage offers more insights into the intrinsic structures of \pmr\, problems. 
\ifdefined\FULLPAPER
\else
A few non-essential proofs are also omitted and can be found in \texttt{http://people.csail.mit.edu/jingjin/files/wafr14.pdf}.
\fi
}
\vspace*{-3mm}
\section{Pebble Motion Problems}\label{sec:formulation}
\vspace*{-3mm}
Let $G = (V, E)$ be a connected undirected graph with $|V| = n$. Let there be a set $p \le n $ pebbles, numbered $1, \ldots, p$, residing on distinct vertices of $G$. A {\em configuration} of these pebbles is a sequence $S = \langle s_1, \ldots, s_p \rangle$, in which $s_i$ denotes the vertex occupied by pebble $i$. A configuration can also be viewed as a bijective map $S: \mathbb \{1, \ldots, p\} \to V(S)$ in which $V(S)$ denotes the set of occupied vertices by $S$. We allow two types of {\em moves} of pebbles. In a {\em simple move}, a pebble may move to an adjacent empty vertex. In a {\em rotation}, pebbles occupying all vertices of a cycle can rotate simultaneously (clockwise or counterclockwise) such that each pebble moves to the vertex previously occupied by its (clockwise or counterclockwise) neighbor. Two configurations $S$ and $S'$ are {\em connected} if there exists a sequence of moves that takes $S$ to $S'$. Let $S$ and $D$ be two pebble configurations on a given graph $G$, the problem of {\em pebble motion on graphs} is defined as follows. 
\begin{problem}[Pebble Motion on Graphs (\pmg)]\label{pm} Given $(G, S, D)$, find a sequence of simple moves that take $S$ to $D$.
\end{problem}

When $G$ is a tree, \pmg\, is also referred to as {\em pebble motion on trees} (\pmt). In this case, an instance is usually written as $I = (T, S, D)$ with $T$ being a tree. When both simple moves and rotations are allowed, the resulting variant is the problem of {\em pebble motion with rotations}. 
\begin{problem}[Pebble Motion with Rotation (\pmr)]\label{pmr} Given $(G, S, D)$, find a sequence of simple moves and rotations that takes $S$ to $D$.
\end{problem}

If $G$ is a tree, then a \pmr\, is simply a \pmt. We note that it may be possible to achieve additional efficiency by allowing multiple simple moves and rotations (along disjoint cycles) to take place concurrently. For example, the configuration in Fig.~\ref{fig:synchro}(a) can be taken to the configuration in Fig.~\ref{fig:synchro}(b) in a single concurrent move. A full  discussion of such moves ({\em i.e.}, the optimality perspective) is beyond the scope of this paper. 

\vspace*{-3mm}
\section{Graph Induced Group and the Upper Bound on its Diameter}\label{sec:upper}
\vspace*{-3mm}
\subsection{Groups Generated by Cyclic Pebble Motions and their Diameters}\label{sub:cyclic} 
\vspace*{-2mm}
A particularly important case of \pmr\, is when $p = n$; we restrict our discussion to this case in this section. When $p = n$, only synchronous rotations are possible. Given two configurations $S$ and $S'$ that are connected, they induce a permutation of the pebbles, which is computable via $\sigma_{S,S'}(i) = S^{-1}(S'(i))$ for each pebble $i$; $\sigma_{S, S}$ is the identity element. Given an initial configuration $S_0$, let $\mathcal S$ denote the set of all configurations reachable from $S_0$. It can be verified, using basic definitions of groups, that the permutations $\sigma_{S_0, S_i}$ over all $S_i \in \mathcal S$ form a subgroup of $\mathbf{S_n}$, the symmetric group on $n$ letters. Since this group is determined by the graph $G$, we denote it $\G$. 
\begin{figure}[htp]
\vspace*{-6mm}
\begin{center}
    \includegraphics[width=0.28\textwidth]{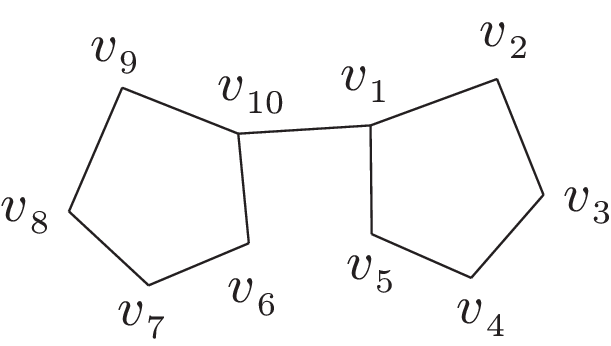} 
\end{center}
\vspace*{-6mm}
\caption{\label{fig:cycles} For the graph above, the collection of sets of cycles are $\mathcal C = \{\{v_1v_2v_3v_4v_5\}, $ $ \{v_6v_7v_8v_9v_{10}\},$ $ \{v_1v_2v_3v_4v_5, v_6v_7v_8v_9v_{10}\} \}$.}
\end{figure}
\vspace*{-5mm}

Two cycles of $G$ are {\em disjoint} if their vertex sets have empty intersection. When $p = n$, each synchronous move corresponds to the rotations of pebbles along a set of of disjoint cycles. Let $\mathcal C$ be the collection of all sets of disjoint cycles in $G$; each $C \in \mathcal C$ is a unique set of disjoint cycles of $G$. Since the pebbles may rotate clockwise or counterclockwise along a cycle $c_i \in C$, each set of disjoint cycles $C$ can take a configuration to $2^{|C|}$ new configurations with one move. That is, each $C$ yields $2^{|C|}$ generators of $\G$. Let the set of all generators obtained this way be $\mathcal G$. As an example, the graph in Fig. \ref{fig:cycles} has two cycles, with $|\mathcal C| = 3$ and $|\mathcal G| = 8$ (note that $|\mathcal G| = 2^{|\mathcal C|}$ does not hold in general). We make the simple observation that these definitions yield a natural bijection between synchronous moves and elements of $\mathcal G$. As such, when a configuration $S'$ is reachable from a configuration $S$, we say that the permutation $\sigma_{S, S'} \in \G$ is {\em reachable} (from the identity) using products of generators from $\mathcal G$ corresponding to the synchronous moves. We frequently invoke this bijection between synchronous moves and generators without explicitly stating so. Lastly, any element $x \in \G$ can be expressed as generator product $g_1g_2\ldots g_k$ in which $g_1, \ldots, g_k \in \mathcal G$. Let $k_x$ be the minimum $k$ such that $x = g_1g_2\ldots g_k$. The diameter of $\G$, $diam(\G)$, is defined as the maximum $k_x$ over all $x \in \G$. 
\vspace*{-3mm}
\subsection{Upper Bound over Group Diameters}
\vspace*{-2mm}
The main result to be established in this section is $diam(\G) = O(n^2)$. To show this, $G$ is divided into classes based on its connectivity. When $G$ is connected (1-connected) but none of its subgraphs are 2-connected ({\em i.e.}, $G$ has no cycles), it is a tree. In this case, no pebble can move. Another simple case is when $G$ is a cycle, the simplest 2-connected graph. Then, it is clear that all elements of $\G$ are generated by a single rotation. 
\begin{lemma}[Trees and Cycles]\label{l:treecycle} If $G$ is a tree, then $\G \cong \{1\}$, the trivial group. If $G$ is a cycle, then $\G \cong \mathbb Z/n$, the cyclic group of order $n$. \end{lemma}
\begin{figure}[htp]
\vspace*{-10mm}
\begin{center}
    \includegraphics[width=0.35\textwidth]{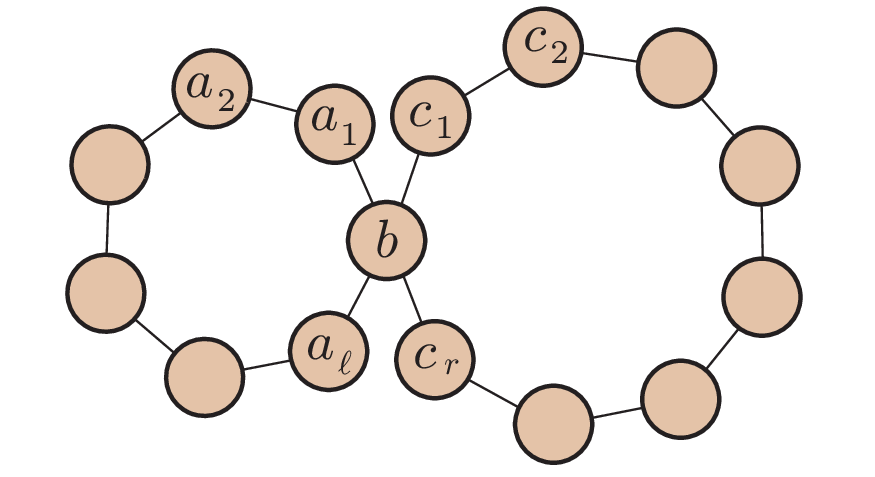} 
\end{center}
\vspace*{-6mm}
\caption{\label{fig:separable}Two cycles sharing one common vertex. The graph is {\em separable} at $b$.}
\end{figure}
\vspace*{-5mm}

When $G$ is connected but the removal of some vertex from $G$ leaves two or more components, it is {\em separable}. An important case here is when $G$ is a set of cycles sharing vertices so that no edge of $G$ is on more than one cycle. Such graphs form a subset of $2$-edge-connected graphs. Fig. \ref{fig:separable} gives an example with two cycles. Following convention, $\mathbf{A_n}$ denotes the {\em alternating group} on $n$ letters. For groups, $\G_1 \ge \G_2$ or $\G_2 \le \G_1$ denotes that $\G_2$ is a subgroup of $\G_1$. For two configurations $S$ and $S'$ over the same set of pebbles on the same graph, we say that they are {\em cycle similar} if the following property holds. For any pebble $a$, let the sets of cycles (of the underlying graph $G$) occupied by $a$ in configurations $S$ and $S'$ be $C_S$ and $C_{S'}$, respectively. Then $C_S \cap C_{S'} \ne \varnothing$. 

A key result of this section is the following. 

\begin{theorem}[Cycles, Separable]\label{t:separable}If every edge of a separable graph $G$ is on exactly one cycle, then $\G \ge \mathbf{A_n}$ and $diam(\G) = O(n^2)$.
\end{theorem}
{\sc Proof.} Given configurations $S$ and $D$, we claim: 

1. In $O(n^2)$ moves, $D$ can be taken to some configuration $D'$ such that $S$ and $D'$ are cycle similar. As an example, in Fig. \ref{fig:separable}, assuming the given configuration is $S$, this step ensures that in configuration $D'$, pebbles $a_i$'s are all on the left cycle and pebbles $c_i$'s are all on the right cycle. The pebble $b$ may appear on either one of the two cycles.  

2. In $O(n^2)$ moves from $D'$, a configuration $D''$ can be reached such that either $D'' = S$ or $D''$ and $S$ differ by a transposition (group action). We require that the transposition is fixed for a fixed $S$ and involves two adjacent pebbles of $S$. Let $S'$ be the result of letting this transposition act on $S$.  

These claims are proved in lemmas that follow. By these claims, an arbitrary $D$ can reach either $S$ or $S'$. Therefore, all configurations (and consequently elements of $\mathbf{S_n}$) are partitioned into two equivalence classes based on mutual reachability. Since the only subgroup of $\mathbf{S_n}$ of index 2 is $\mathbf{A_n}$, this implies that $\G \ge \mathbf{A_n}$. 

When $\G \cong \mathbf{A_n}$, any element of $\G$ is a product of generators from $\mathcal G$ with a length of $O(n^2)$, proving $diam(\G) = O(n^2)$. If $\G$ is not isomorphic to $\mathbf{A_n}$, since the only subgroups of $\mathbf{S_n}$ containing $\mathbf{A_n}$ are $\mathbf{A_n}$ and $\mathbf{S_n}$ itself, $\G \cong \mathbf{S_n}$. This implies that $\mathbf{A_n}$ has at most two cosets in $\G$; denote the other coset of $\mathbf{A_n}$ as $\mathbf{A_n}^c$, which also have a diameter of $O(n^2)$ (to see this, note that any configuration $D$ is reachable from one of $S$, $S'$ in $O(n^2)$ moves). From the identity, all elements of $\mathbf{A_n}$ are reachable using generator products of length $O(n^2)$. Since elements of $\mathbf{A_n}^c$ are now reachable from elements of $\mathbf{A_n}$, an element of $\mathbf{A_n}^c$ must be reachable from the identity using a generator product of length $O(n^2)$ as well. Therefore, when $\G \cong \mathbf{S_n}$, all elements of $\G$ are reachable using generator products of length $O(n^2)$, yielding $diam(\G) = O(n^2)$. ~\qed

\begin{figure}[htp]
\vspace*{-6mm}
\begin{center}
    \includegraphics[width=0.45\textwidth]{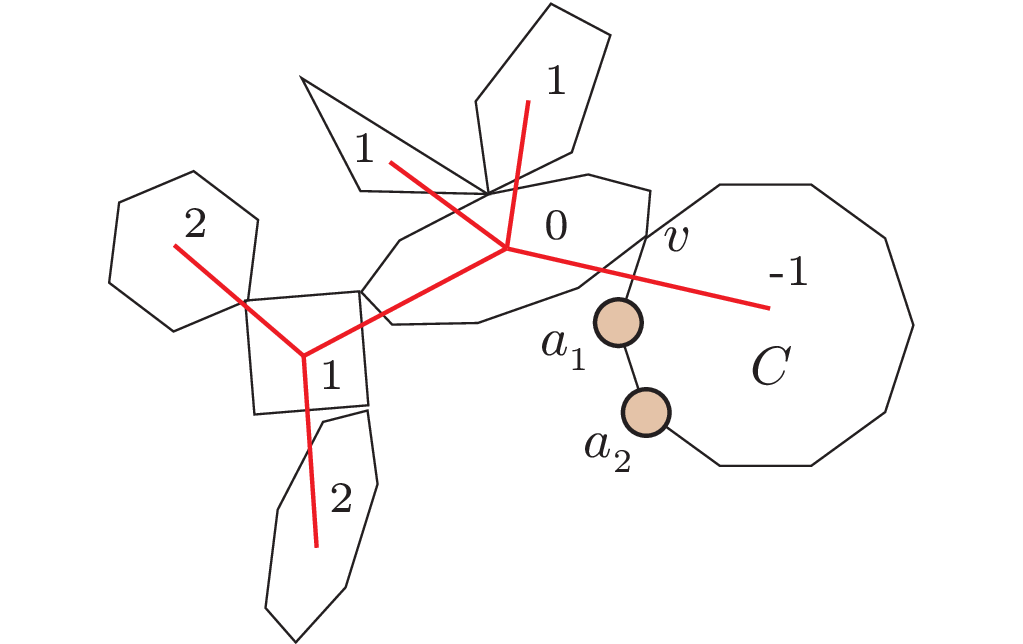} 
\end{center}
\vspace*{-6mm}
\caption{\label{fig:cycle-tree} The dual tree structure in a separable graph $G$ with every edge on exactly one cycle. The numbers represent the cycle distances of the cycles to the leaf cycle $C$, which in fact is the root of the tree.}
\end{figure}

Before moving to the lemmas, we note that when $G$ is separable and every edge of $G$ is on exactly one cycle, the edges of $G$ can be partitioned into equivalence classes based on the cycles they belong to. Because $G$ is separable, every cycle must border one or more cycles and at the same time, two cycles can share at most one vertex. Such a graph is also called a {\em cactus} graph. Moreover, there exists a cycle that only shares one vertex with other cycles. We call such a cycle a {\em leaf cycle}. An example of a leaf cycle is given in Fig. \ref{fig:cycle-tree}.
\vspace*{-5mm}

\begin{figure}[htp]
\vspace*{-5mm}
\begin{center}
    \includegraphics[width=0.6\textwidth]{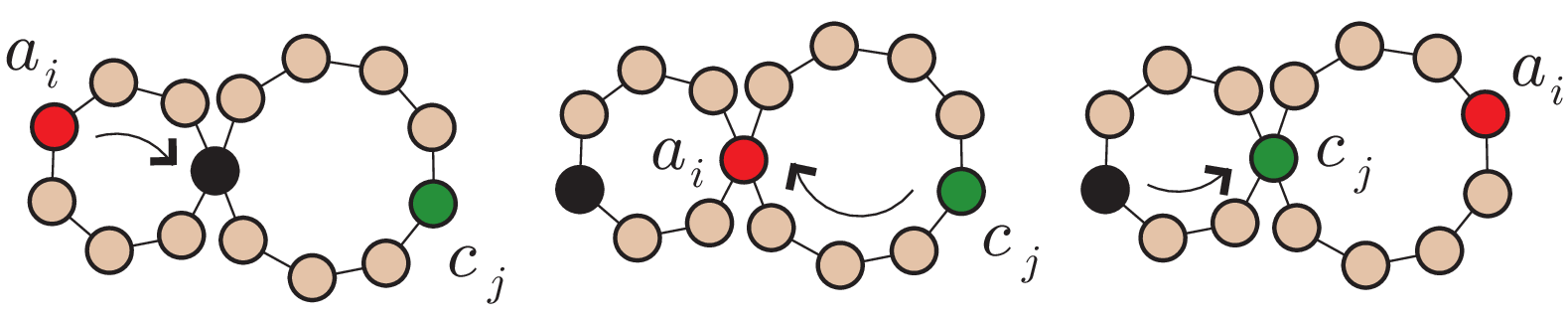} 
\end{center}
\vspace*{-5mm}
\caption{\label{fig:arrange}Illustration of the vertex arrange algorithm for two adjacent cycles.}
\end{figure}

Given a cycle $C'$ on $G$, it is of {\em cycle distance} $d_c$ to $C$ if a vertex on $C'$ needs to travel through at least $d_c$ cycles to reach $C$. A neighboring cycle of $C$ has distance 0 since they share a common vertex. Let $C$ have a cycle distance of $-1$ by definition. This induces a (dual) tree structure on the cycles when viewing them as vertices joined by edges to neighbors (see, {\em e.g.}, Fig. \ref{fig:cycle-tree}). Computing such a tree takes time $O(|V| + |E|)$ because obtaining maximal $2$-connected components takes linear time \cite{Tar72}. The first claim in the proof of Theorem \ref{t:separable} can be stated as follows. 
\begin{lemma}[Initial Arrangement]\label{l:arrange} Given a separable $G$ with each edge on exactly one cycle and configurations $S$ and $D$, in $O(n^2)$ moves, a configuration that is cycle similar to $S$ is reachable from $D$.
\end{lemma}
{\sc Proof.} Note that a pebble may reside on multiple cycles; this lemma only ensures that each pebble gets moved to one of the cycles it belongs to in $S$. First we show that a single pebble can be relocated to a cycle it belongs to in $S$ in $O(n)$ rotations, without affecting pebbles that are previously arranged. When $G$ is two cycles joined on a common vertex ({\em e.g.}, Fig. \ref{fig:separable}), without loss of generality, assume that we need to move $a_i$ from the left cycle to the right cycle. This implies that some pebble $c_j$ (and possibly $b$) does not belong to the right cycle in $S$. We note that the group $\G$ in this case has four generators, 
$
g_{\ell} = \left(
\begin{array}{lllll}
a_1 & a_2 & \ldots & a_{\ell} & b \\
b & a_1 & \ldots & a_{\ell-1} & a_{\ell}
\end{array}\right), 
g_{r} = \left(
\begin{array}{lllll}
c_1 & c_2 & \ldots & c_{r} & b \\
c_2 & c_3 & \ldots & b & c_1
\end{array}\right), 
$
which correspond to clockwise rotations along the left and right cycles, respectively, and their inverses, $g_{\ell}^{-1}$ and $g_r^{-1}$. One can verify that the generator product $g_{\ell}^{-i}g_r^{-j}g_{\ell}^i$ exchanges $a_i$ and $c_j$ between the two cycles without affecting the cycle membership of other pebbles (see Fig. \ref{fig:arrange}).
For the general case in which a pebble needs to go through some $k$ cycles, denoting the generators as $g_1, \ldots, g_k$, it is easy to verify that a product of the form $g_1^{-i_1}g_2^{-i_2}\ldots g_k^{i_k}\ldots g_2^{i_2}g_1^{i_1}$ achieves what we need, with $i_1 + \ldots + i_k < n$. There may be more than these $2k$ basic generators, but we do not need the other generators for this proof. Therefore, at most $2n$ moves are needed to move one pebble to the desired cycle. To avoid affecting pebbles that are previously arranged, we may simply fix a leaf cycle $C$ and start with cycles  based on their cycle distance to $C$ in decreasing order. At most $2n^2$ moves are required to arrange all $n$ pebbles to the desired cycles. ~\qed
\begin{lemma}[Rearrangement]\label{l:rearrange}The pebbles arranged according to Lemma \ref{l:arrange} can be rearranged such that the resulting configuration is the same as $S$ or differ from $S$ by a fixed transposition of two neighboring pebbles in $S$. Rearrangement requires $O(n^2)$ moves. 
\end{lemma}
{\sc Proof.} For a fixed $G$, let $C$ be a leaf cycle and let $C$ border other cycle(s) via vertex $v$. In $S$, let $a_1$ be the pebble occupying counterclockwise neighboring vertex of $v$ on the cycle $C$, and let $a_2$ be the counterclockwise neighbor of $a_1$ on $C$ (again, see Fig. \ref{fig:cycle-tree} for an illustration of this setup). The fixed transposition will be $(a_1 \, a_2)$. 

\begin{figure}[htp]
\vspace*{-5mm}
\begin{center}
    \includegraphics[width=0.6\textwidth]{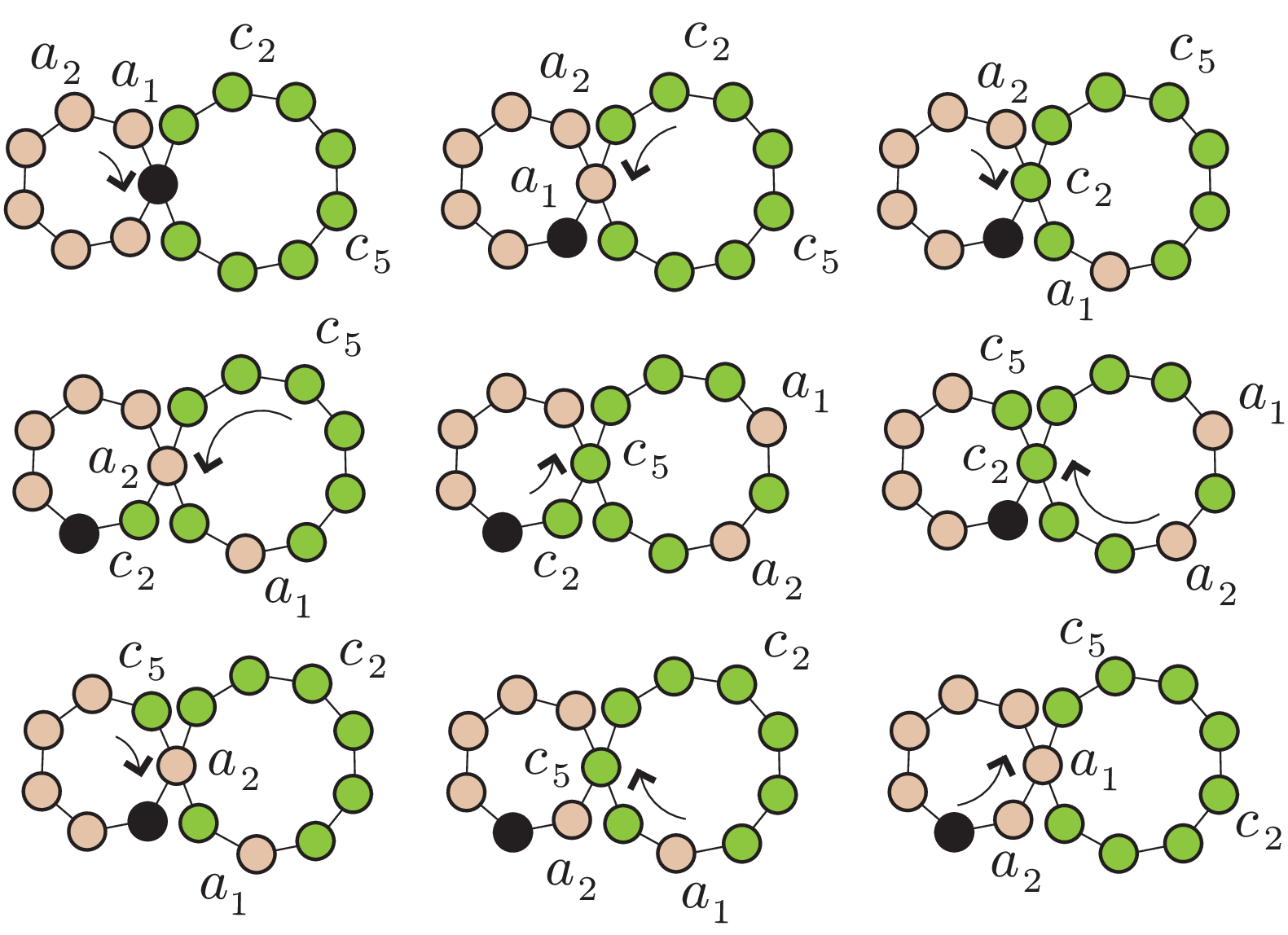} 
\end{center}
\vspace*{-5mm}
\caption{\label{fig:cycle-exchange}Illustration of the rearrangement algorithm (from left to right, then top to bottom).}
\end{figure}

We rearrange pebbles to match the configuration $S$ starting from cycles with higher cycle distances to the leaf cycle $C$, using the neighboring cycle with smaller cycle distance (such a cycle is unique). We show that the pebbles on the more distant cycle can always be rearranged to occupy the vertex specified by $S$. Moreover, this can be achieved using moves that only affect the ordering of two pebbles on the neighboring cycle. Without loss of generality, we use the two cycle example from Fig. \ref{fig:separable} and let the right cycle be the more distant one. The generators $g_{\ell}, g_{\ell}^{-1}$, $g_r$, and $g_r^{-1}$ from previous lemma remain the same. To exchange two pebbles on the right cycle, for example $c_i, c_j$, we may use the following generator product
\begin{equation}\label{eq:seq} g_{\ell}^{-2}g_r^{-i}g_{\ell}g_r^{j - i}g_{\ell}^{-1}g_r^{-j + i}g_{\ell}g_r^{-i}g_{\ell}.
\end{equation}
It is straightforward to verify that (\ref{eq:seq}) works. To make it clear, Fig. \ref{fig:cycle-exchange} illustrates the application of (\ref{eq:seq}) for exchanging $c_2$ and $c_5$ using $a_1, a_2$. Every such exchange requires at most $2n$ moves.

Performing such exchanges iteratively, within $2n^2$ moves, all pebbles except those on the leaf cycle $C$ can be rearranged to occupy vertices specified by $S$. Reversing the process, we can arrange all pebbles on $C$ to occupy vertices specified by $S$, using a neighboring cycle $C'$, affecting the ordering of at most two pebbles on $C'$. Repeating this process again with $C'$ using $C$ as the neighboring cycle and $a_1, a_2$ as the swapping pebbles, all pebbles except possibly $a_1, a_2$ occupy the vertices specified by $S$. ~\qed

The above two lemmas complete the proof of Theorem~\ref{t:separable}. At this point, it is easy to see that when $G$ is separable with each edge on a single cycle, $\G \cong \mathbf{S_n}$ if and only if $G$ contains an even cycle, corresponding to the composition of an odd number of transpositions. Otherwise, $\G \cong \mathbf{A_n}$. We are left with the case in which $G$ is 2-connected but not a (single) cycle. 

\begin{theorem}[2-connected, General]\label{t:2-switch}If $G$ is 2-connected and not a cycle, $\G \cong \mathbf{S_n}$ with $diam(\G) = O(n^2)$. 
\end{theorem}
\ifdefined\FULLPAPER
{\sc Proof.} Our proof again starts by showing that the locations of two pebbles can be exchanged without affecting the locations of other pebbles. Given a 2-connected graph $G$ that is not a cycle, it can always be decomposed into a cycle plus one or more {\em ears} (an ear is a simple path $P$ whose two end points lie on some cycle that does not contain other vertices of $P$). Therefore, any two pebbles on $G$ must lie on some common cycle with one attached ear. We may then assume that the two pebbles to be exchanged lie somewhere on two adjacent cycles ({\em i.e.}, they are two arbitrary pebbles in Fig. \ref{fig:three-cycle}). Restricting to such a graph $G'$ of $G$, which has three cycles (left, right, and outer), rotations along these cycles will not affect the rest of the pebbles not on $G'$. We claim that moving within $G'$ is sufficient to exchange any two pebbles on $G'$ and the operation can be done with $O(n)$ moves. 
\begin{figure}[htp]
\vspace*{-5mm}
\begin{center}
    \includegraphics[width=0.35\textwidth]{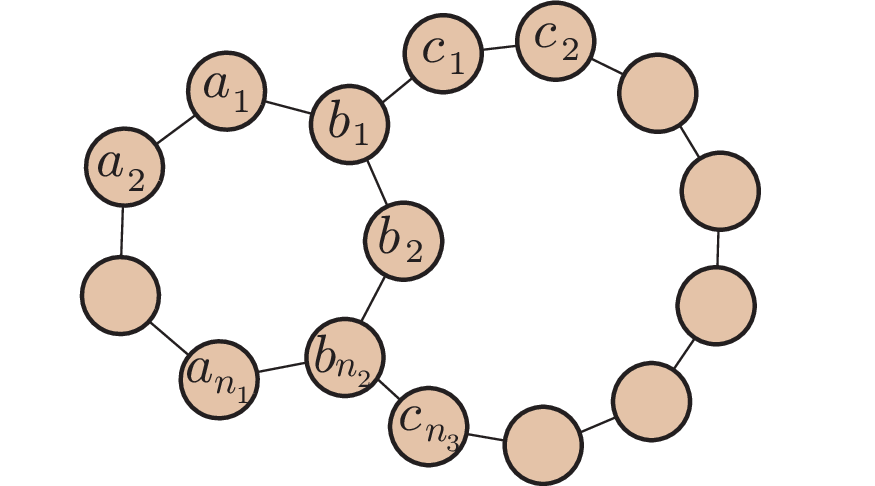} 
\end{center}
\vspace*{-5mm}
\caption{\label{fig:three-cycle} A simple 2-connected graph. There are six moves for this configuration: Rotating clockwise or counterclockwise along one of the three cycles.}
\end{figure}

Let $G'$ have $n_1 + n_2 + n_3$ vertices, with $n_1$ vertices belonging to the left cycle only, $n_3$ vertices belonging to the right cycle only and $n_2$ vertices shared by the two cycles. Assuming the initial pebble configuration is as illustrated in Fig. \ref{fig:three-cycle}, we have the following generators,
\begin{displaymath}
\begin{array}{l}
g_{\ell} = \left(
\begin{array}{lllllll}
a_1 & a_2 & \ldots & a_{n_1} & b_{n_2} & \ldots & b_1 \\
b_1 & a_1 & \ldots & a_{n_1-1} & a_{n_1} & \ldots & b_2
\end{array}\right), \\
g_{r} = \left(
\begin{array}{lllllll}
c_1 & c_2 & \ldots & c_{n_3} & b_{n_2} & \ldots & b_1 \\
c_2 & c_3 & \ldots & b_{n_2} & b_{n_2-1} & \ldots & c_1
\end{array}\right), \\
g_{o} = \left(
\begin{array}{llllllll}
b_1 & c_1 & \ldots & c_{n_3} & b_{n_2} & a_{n_1} & \ldots & a_1 \\
c_1 & c_2 & \ldots & b_{n_2} & a_{n_1} & a_{n_1-1} & \ldots & b_1
\end{array}\right), 
\end{array}
\end{displaymath} 
which are clockwise rotations along the left, right, and the outer cycles of $G'$, and their inverses,  $g_{\ell}^{-1}, g_r^{-1}$, and $g_o^{-1}$. Note that  
\begin{equation}\label{eq:3}
g_rg_{\ell}g_o^{-1} = 
\left(
\begin{array}{ll}
b_1 & c_1 \\
c_1 & b_1 
\end{array}\right) = (b_1 \, c_1). 
\end{equation}
That is, we may exchange (transpose) $b_1$ and $c_1$ using a generator product of length $3$. Using this length 3 product $g_rg_{\ell}g_o^{-1}$, it is possible to exchange any two pebbles on $G'$ without affecting other pebbles. We elaborate two such cases, all other cases are similar. In a first case we exchange $a_i$ and $c_j$. To do this, we first move $c_j$ to $c_1$'s location, followed by moving $a_i$ to $b_1$'s location. We can then switch $a_i$ and $c_j$ using the primitive $g_rg_{\ell}g_o^{-1}$. Reversing the earlier steps then switches $a_i$ and $c_j$ without affecting any other pebbles. The complete product sequence is $g_{\ell}^{-i}g_r^jg_{\ell}g_o^{-1}g_r^{-j+1}g_{\ell}^i$, which requires $O(n)$ moves or generator actions. Similarly, if we want to switch some $c_i, c_j$ that are not adjacent, we can move them along the outer cycle until one of them belongs to the left cycle and the other to the right cycle. The case of exchanging $a_i, c_j$ then applies, after which we reverse the earlier moves on the outer cycle to obtain the net effect of switching $c_i, c_j$. The number of moves is again $O(n)$. This implies $\G \cong \mathbf{S_n}$ and $diam(\G) = O(n^2)$. 
~\qed
\else
We omit the proof of Theorem~\ref{t:2-switch} but mention that it has a similar general structure as the proof of Theorem~\ref{t:separable}.
\fi
Combining Theorems \ref{t:separable} and \ref{t:2-switch} concludes the case for $2$-edge-connected graphs that are not single cycles; the case of general graph then follows. Since we will mention ``2-edge-connected component'' fairly frequently, we abbreviate it to ``TECC'' except in theorem statements. Also, we call each component of $G$ after deleting all TECCs a {\em branch}. 
\begin{proposition}[$2$-edge-connected]\label{p:2-edge} If $G$ is $2$-edge-connected and not a single cycle, $\G \ge \mathbf{A_n}$ with $diam(\G) = O(n^2)$. 
\end{proposition}
{\sc Proof.} A $2$-edge-connected graph $G$ can be separated into $2$-connected components via splitting at articulating vertices. A (dual) tree structure, similar to that illustrated in Fig. \ref{fig:cycle-tree}, can be built over these components. The two-step algorithm used in the proof of Theorem \ref{t:separable}, in combination with Theorem~\ref{t:2-switch}, can be applied to show that $\G \ge \mathbf{A_n}$ and $diam(\G) = O(n^2)$. ~\qed

\vspace{2mm}
After gathering all cases, we obtain the following main result for this section. 
\begin{theorem}[General Graph]\label{t:general} Given an arbitrary connected, undirected, simple graph $G$, $diam(\G) = O(n^2)$.
\end{theorem}
{\sc Proof.} Pebbles on vertices of $G$ that are not on any cycle are always immobile. Deleting those vertices does not change $\G$. After all such vertices are removed, we are left with the TECCs of $G$. Denoting the associated groups of these components $\{\G_i\}$, $\G$ is the direct product of the $\G_i$'s. Since all $\G_i$'s have $O(n^2)$ diameter, so does $\G$.~\qed

\vspace*{-4mm}
\section{Linear Time Feasibility Test of \pmr}\label{sec:linear}
We now describe a linear time algorithm for testing the feasibility for \pmr, using a proof strategy similar to that from \cite{AulMonParPer99} on \pmt. We first restate a result form \cite{AulMonParPer99}.
\begin{theorem}[Theorem 3 in \cite{AulMonParPer99}]\label{t:pmt-ppt} Given an instance $(T, S, D)$ of \pmt, in $O(n)$ steps, an instance $(T, S', D)$ of \pmt\, can be computed such that $S'$, $D$ contain the same set of vertices and $(T, S, S')$ is feasible.
\end{theorem}

The following corollary is also obvious. 
\begin{corollary}\label{c:pmt}Given an instance $(T, S, D)$ of \pmr, let $(T, S', D)$ be the new instance obtained according to Theorem \ref{t:pmt-ppt}. Then $(T, S, D)$ is feasible if and only if $(T, S', D)$ is feasible.
\end{corollary}

\def\ppr{PPR}
\def\rpp{RPP}
By Theorem~\ref{t:pmt-ppt} and Corollary~\ref{c:pmt}, reconfiguration can be performed on a \pmr\, instance $I = (G, S, D)$ to get an equivalent instance $I' = (G, S', D)$ so that $S', D$ have the same underlying vertex set ({\em i.e.}, $V(S') = V(D)$). To do this, find a spanning tree $T$ of $G$. The $O(n)$ time algorithm guaranteed by Theorem \ref{t:pmt-ppt} can then compute a desired instance $(T, S', D)$ with $S', D$ having the same set of vertices. Since the moves taking $(T, S, S')$ is feasible, $(G, S, S')$ is feasible; therefore, $(G, S, D)$ is feasible if and only if $(G, S', D)$ is feasible. Given an instance $I = (G, S, D)$ in which $S$ and $D$ have the same underlying set, we call it the {\em pebble permutation with rotation} problem or \ppr. Given a \ppr\, instance, we say that two pebbles are {\em equivalent} if they can exchange locations with no net effect on the locations of other pebbles. A set of pebbles are equivalent if every pair of pebbles from the set are equivalent. 

In testing the feasibility of a \ppr\, instance $I = (G, S, D)$, a simple but special case is when $G$ is a cycle. In this case, $S$ and $D$ induce natural cyclic orderings of the pebbles. The following is then clear. 
\begin{lemma}\label{o:cycle} Let $I = (G, S, D)$ be an instance of \ppr\, in which $G$ is a cycle. Then $I$ is feasible if and only if $s_i = d_{(i + k) \textrm{ mod } p}$ for some fixed natural number $k$.
\end{lemma}

When $G$ is not a cycle, the feasibility test is partitioned into four main cases, depending on the number of pebbles, $p$, with respect to the number of vertices of $G$. It is assumed that $G$ contains at least one TECC since otherwise $G$ is a tree and the problem is a \pmt\, problem. 

\vspace*{-4mm}
\subsection{Feasibility test of \ppr\, when $p = n$}
When $p = n$, all vertices are occupied by pebbles. Clearly, if a pebble is on a vertex that does not belong to any cycle ({\em i.e.}, a branch vertex), the pebble cannot move. Therefore, $I = (G, S, D)$ is feasible only if for every branch vertex $v \in V(G)$, $S^{-1}(v) = D^{-1}(v)$. Furthermore, given any TECC $C$ of $G$, $S^{-1}(C) = D^{-1}(C)$ must also hold, since pebbles cannot move out a TECC. If these conditions hold, the feasibility of $I$ is reduced to feasibilities of $\{(C_i, S |_{S^{-1}(C_i)}, D |_{D^{-1}(C_i)})\}$, in which $C_i$'s are the TECCs of $G$ and $S |_{S^{-1}(C_i)}$ denotes $S$ restricted to the domain $S^{-1}(C_i)$; same applies to $D |_{D^{-1}(C_i)}$. More formally,
\begin{proposition}\label{p:peqn} Let $I = (G, S, D)$ be an instance of \ppr\, with $p = n$. Let $\{C_i\}$ be the set of 2-edge-connected components of $G$. Then $I$ is feasible if and only if the following holds: 1. for all $v \in V(G \backslash (\cup_i C_i))$, $S^{-1}(v) = D^{-1}(v)$, 2. for each $C_i$, $S^{-1}(C_i) = D^{-1}(C_i)$, and 3. for each $C_i$, the \ppr\, instance $(C_i, S |_{S^{-1}(C_i)}, D |_{D^{-1}(C_i)})$ is feasible. Moreover, the feasibility test can be performed in linear time. 
\end{proposition}

\noindent {\sc Proof.} Finding TECCs of $G$ can be done in $O(|V| + |E|)$ time \cite{Tar72}. Checking whether condition 1 holds takes linear time. For checking condition 2, for each $C_i$, we first gather $S^{-1}(C_i)$ and for each pebble in $S^{-1}(C_i)$, mark the pebble as belonging to $C_i$. We can then check whether the pebbles in $D^{-1}(C_i)$ also belong to $C_i$ in linear time. For condition 3, deciding the feasibility of $(C_i, S |_{S^{-1}(C_i)},D |_{D^{-1}(C_i)})$ can be done using the results from Section \ref{sec:upper}. This check can performed as follows. 1. Check whether $C_i$ is a cycle, which is true if and only if no vertex of $C_i$ has degree more than two. If this is the case, apply Observation \ref{o:cycle} to test the feasibility on $C_i$; 2. Check whether $C_i$ is a cactus with no even cycle. We can verify whether $C_i$ is a cactus as follows: Using depth first search (DFS), detecting cycles of $C_i$. If $C_i$ is a cactus, then it should assume a ``tree'' structure shown in Fig. \ref{fig:cycle-tree}; the first cycle that is found must be a leaf cycle. Deleting this cycle (without deleting the vertex that joins this cycle to the rest of $C_i$) from $C_i$ yields another cactus. Repeating the process tells us whether $C_i$ is a cactus. As we are finding the cycles, we can check whether there is an even cycle. If $C_i$ is indeed a cactus with no even cycle, the possible configurations have two equivalence classes. The subproblem is only infeasible if $S |_{S^{-1}(C_i)}, D |_{D^{-1}(C_i)}$ fall into different equivalence classes, which can be checked by computing the parity of the permutation $\sigma_{S, D}$, restricted to $C_i$, in linear time; 3. For all other types of $C_i$, the subproblem is feasible. ~\qed

\subsection{Feasibility test of \ppr\, when $p = n - 1$ }
When $p = n - 1$, nearly all \ppr\, instances, in which $G$ are 2-edge-connected graphs, are feasible. 
\begin{lemma}\label{l:tec-feasible} Let $I = (G, S, D)$ be an instance of \ppr\, in which $G$ is 2-edge-connected and not a cycle. If $p < n$, then $I$ is feasible.  
\end{lemma}
\noindent {\sc Proof.} By Theorems \ref{t:separable} and \ref{t:2-switch}, $\G \ge \mathbf{A_n}$. That is, there are at most two equivalence classes of configurations, with configurations from different classes differ by a transposition of neighboring pebbles. Since there is at least one empty vertex, viewing that vertex as a ``virtual'' pebble that can be exchanged with a neighboring pebble in one move, it is then clear that the two configuration classes collapse into a single class. ~\qed
\begin{lemma}\label{l:spoke} Let $I = (G, S, D)$ be an instance of \ppr\, in which $G$, after deleting one (or more) degree $1$ vertex (vertices), is a 2-edge-connected graph. If $p < n$, then $I$ is feasible. 
\end{lemma}
\noindent {\sc Proof.} Note that by degree 1 vertices, we mean that these vertices have degree 1 in $G$. Let $H$ be the 2-edge-connected graph after deleting all degree 1 vertices and let ${v_1, \ldots, v_k}$ be the degree 1 vertices. Let the neighbor of $v_i$ in $G$ be $v_i' \in V(H)$. Since $v \in {v_1, \ldots, v_k}$ has degree 1, it is attached to $H$ via a single edge. Let $H_i$ be the subgraph of $G$ after deleting all vertices in ${v_1, \ldots, v_k}$ except $v_i$. Assume that $v_1$ is empty initially, we show next that all pebbles occupying $H_1$ are equivalent. That is, an arbitrary configuration of these pebbles can be achieved. 
\begin{figure}[htp]
\vspace*{-5mm}
\begin{center}
    \includegraphics[width=0.6\textwidth]{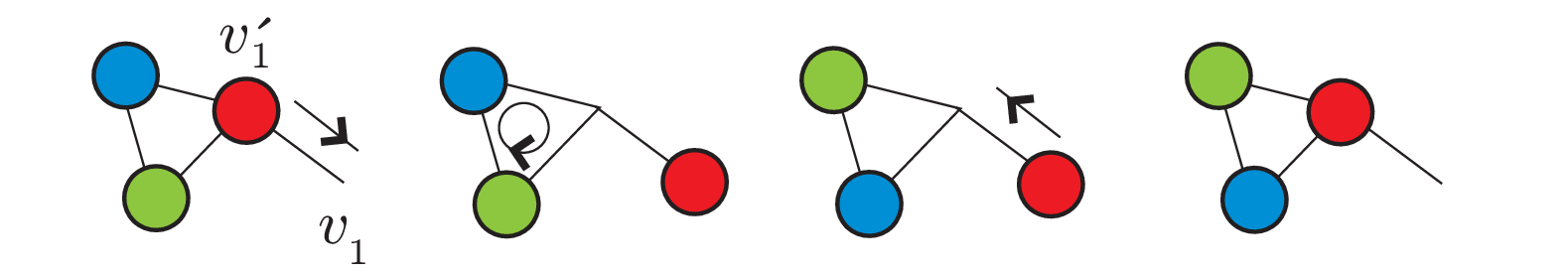} 
\end{center}
\vspace*{-5mm}
\caption{\label{fig:vacant-cycle} With one empty vertex, pebbles on a triangle can be arranged to achieve any desired configuration. This generalizes to an arbitrary TECC.}
\end{figure}

If $H$ is cycle, the subroutine illustrated in Fig \ref{fig:vacant-cycle} shows how an arbitrary configuration of pebbles can be achieved for a triangle $H$, which directly generalizes to an arbitrary sized cycle. This shows that all pebbles on $H_1$ fall in the same equivalence class. If $H$ is not a cycle, we can move an arbitrary pebble $j$ from $H$ to $v_1$. Lemma \ref{l:tec-feasible} implies that all pebbles on $H$ are equivalent. Since $j$ is arbitrary, all pebbles on $H_1$ are equivalent. 

Having shown that all pebbles on $H_1$ are equivalent, we move an arbitrary pebble $j$ to $v_1$ and empty vertex $v_2$ (if there is a $v_2$). Following the same procedure, all pebbles on $H_2$ are equivalent. Since $j$ is arbitrary, all pebbles on $H, v_1, v_2$ are equivalent. Inductively, all pebbles on $G$ are equivalent. Therefore, an arbitrary instance $I$ is feasible. ~\qed

When there is a single empty vertex on $G$, it is clear that pebbles can be moved so that the empty vertex is an arbitrary vertex of $G$. In particular, for any TECC $H$ of $G$, we can move the pebbles so that a vertex of $H$ is empty. By Lemma \ref{l:spoke}, all pebbles on $H$ and its distance one neighboring vertices fall in the same equivalence class. We now show that the feasibility of the case of $p = n - 1$ can be decided in linear time. 
\begin{proposition}\label{p:peqn-1} Let $I = (G, S, D)$ be an instance of \ppr\, in which $p = n -1$ and $G$ is not a cycle. The feasibility of $I$ can be decided in linear time. 
\end{proposition}
\noindent {\sc Proof.} We start with pebble configuration $S$ and group the pebbles into equivalence classes. Without loss of generality, assume that $S$ leaves a vertex of a TECC, say $H$, unoccupied. By Lemma \ref{l:spoke}, all pebbles on $H$ and its distance 1 neighbors belong to the same equivalence class, say $h_{S,1}$. Now, check whether any pebble in $h_{S, 1}$ is on some other TECC $H' \ne H$. If that is the case, all pebbles on $H'$ and its distance 1 neighbors are also equivalent and belong to $h_{S, 1}$. When no more pebbles can be added to $h_{S, 1}$ this way, $h_{S, 1}$ is completely defined. 

Let $v$ be a vertex neighboring a vertex occupied by a pebble from $h_{S, 1}$ ($v$ itself is not occupied by a pebble in $h_{S, 1}$), if $v$ is not a TECC vertex, the pebble currently on $v$ cannot be move to a TECC and therefore is not equivalent to any other pebble. The pebble then gets its own equivalence class, say $h_{S, 2}$. If $v$ belongs to a TECC, say $H_v$, then all pebbles on $H_v$ and all $H_v$'s distance 1 neighbors that are not yet classified belong to $h_{S, 2}$; $h_{S, 2}$ is then expanded similarly to $h_{S, 1}$. At this point, the procedures given so far apply to partition all pebbles into equivalence classes. It is not hard to see the algorithm takes linear time to complete using breadth first or depth first search, treating each TECC as a whole. 
As the start configuration $S$ is being classified, the same is done to $D$. In particular, if a set of pebbles of $S$ belongs to an equivalence class $h_{S, i}$, then the pebbles of $D$ occupying the same set of vertices get assigned to the class $h_{D, i}$. The instance $I$ is feasible if and only if $h_{S, i} = h_{D, i}$ for all $i$ (this can be done in linear time as we have shown in checking the second condition in Proposition \ref{p:peqn}). ~\qed
\begin{figure}[htp]
\vspace*{-5mm}
\begin{center}
    \includegraphics[width=0.4\textwidth]{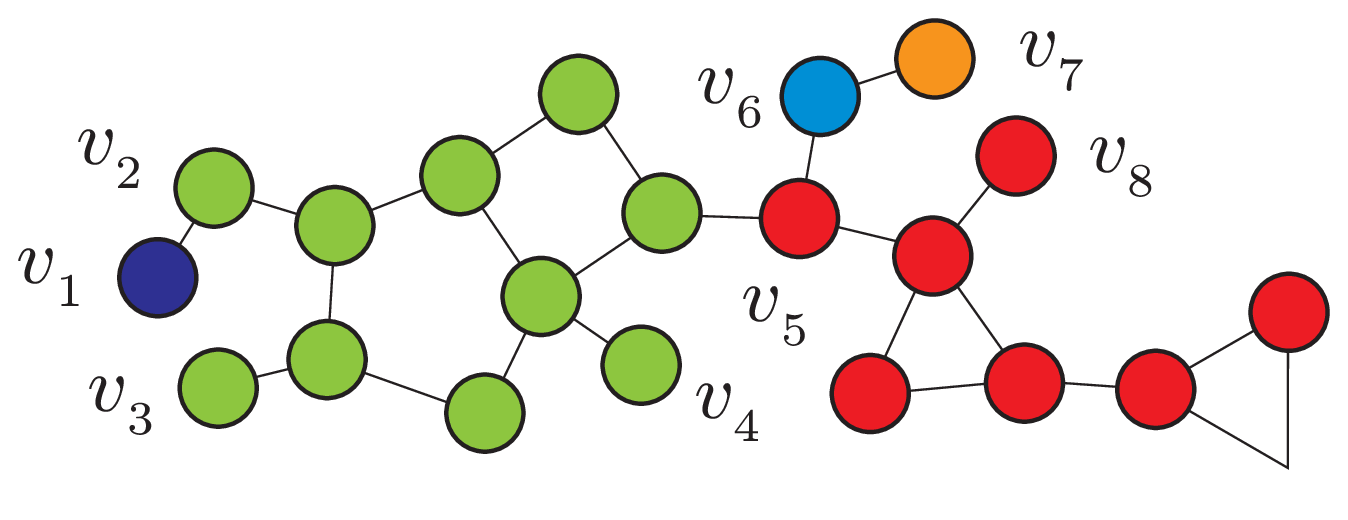} 
\end{center}
\vspace*{-5mm}
\caption{\label{fig:peqn-1} An example of the case $p = n - 1$. The pebbles are put into 5 different equivalence classes, distinguished by different colors.}
\end{figure}
Fig. \ref{fig:peqn-1} provides an example of applying the above procedure to a given pebble configuration, which partitions the pebbles into 5 equivalence classes. 
\vspace*{-3mm}
\subsection{Feasibility test of \ppr\, when $p < N(TECCs)$ }
We denote by $N(TECCs)$ the number of vertices of all TECCs of $G$. An instance is almost always feasible when $p < N(TECCs)$. 

\begin{theorem}\label{t:pln} Let $I = (G, S, D)$ be an instance of \ppr\, in which $G$ is not a cycle. If $p < N(TECCs)$, then $I$ is feasible.  
\end{theorem}
\noindent {\sc Proof.} Since the number of pebbles are not enough to occupy all TECC vertices, we can update configuration $S$ to a new one $S'$ such that all pebbles are on TECC vertices. Repeating the same moves over the configuration $D$ to get $D'$ ({\em i.e.}, if we move a pebble from $v_i$ to $v_j$ in the initial pebble configuration, we move the corresponding pebble from $v_i$ to $v_j$ in the final pebble configuration). After this process is complete, the updated start and final configurations again occupy the same set of vertices; $(G, S, D)$ is feasible if and only if the $(G, S', D')$ is feasible. In the rest of the proof we show that $(G, S', D')$ is feasible. 
\begin{figure}[htp]
\vspace*{-5mm}
\begin{center}
    \includegraphics[width=0.22\textwidth]{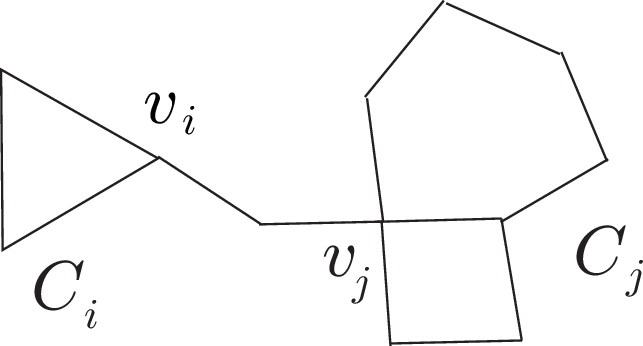} 
\end{center}
\vspace*{-5mm}
\caption{\label{fig:2tec} A graph with two TECCs.}
\end{figure}
\vspace*{-5mm}

Since not all TECC vertices are occupied in $S'$, at least one TECC, say $C_i$, has an empty vertex. By Lamma \ref{l:spoke}, all pebbles on $C_i$ are equivalent. Now let $C_j$ be another TECC joined to $C_i$ via a single branch (see Fig. \ref{fig:2tec} for an example). Since any pebble on $C_j$ can be moved to vertex $v_j$ via a proper sequence of rotations, it is then possible to exchange any pair of pebbles $p_1$ on $C_i$ and $p_2$ on $C_j$: move $p_2$ to $v_j$, empty $v_i$, move $p_2$ to $v_i$, rotate $p_1$ to $v_i$, and move it to $v_j$. Via induction, any pair of pebbles on $G$ can be exchanged, without affecting the current configuration of other pebbles. Given this procedure, we can iteratively arrange each pebble $i$, starting from pebble $1$, by exchanging pebble $i$ with some other pebble occupying $i$'s vertex in $D'$. With up to $p - 1$ exchanges, all pebbles can be arranged to their desired final configurations. ~\qed

\vspace*{-3mm}
\subsection{Feasibility test of \ppr\, when $N(TECCs) \le p < n - 1$ }\label{subsec:rpp}
\vspace*{-3mm}
For this last case, given a \ppr\, instance, $(G, S, D)$, we first move pebbles in $S$ and $D$ so that vertices of all TECCs are occupied. To perform this in linear time, a ``fake'' goal configuration $D_f$ is created with $p$ pebbles such that all TECCs are full occupied, in an arbitrary order. This is possible because $N(TECCs) \le p < n - 1$. Using a spanning tree $T$ of $G$ and apply Theorem \ref{t:pmt-ppt} to $(T, S, D_f), (T, D, D_f)$, we get two new instances $(T, S', D_f)$, $(T, D', D_f)$ with the property that $S', D'$, and $D_f$ all occupy the same set of vertices and $(T, S, S')$, $(T, D, D')$ are both feasible. Thus, we obtain a new \ppr\, instance $(G, S', D')$, which is feasible if and only if $(G, S, D)$ is, with the additional property that vertices of all TECCs are occupied. For convenience, we call an instance $(G, S, D)$ of \ppr\, in which all TECC vertices are occupied a {\em rearranged pebble permutation} problem, or \rpp. Note that this implies $p \ge N(TECCs)$.
\begin{figure}[htp]
\vspace*{-2mm}
\begin{center}
    \includegraphics[width=0.4\textwidth]{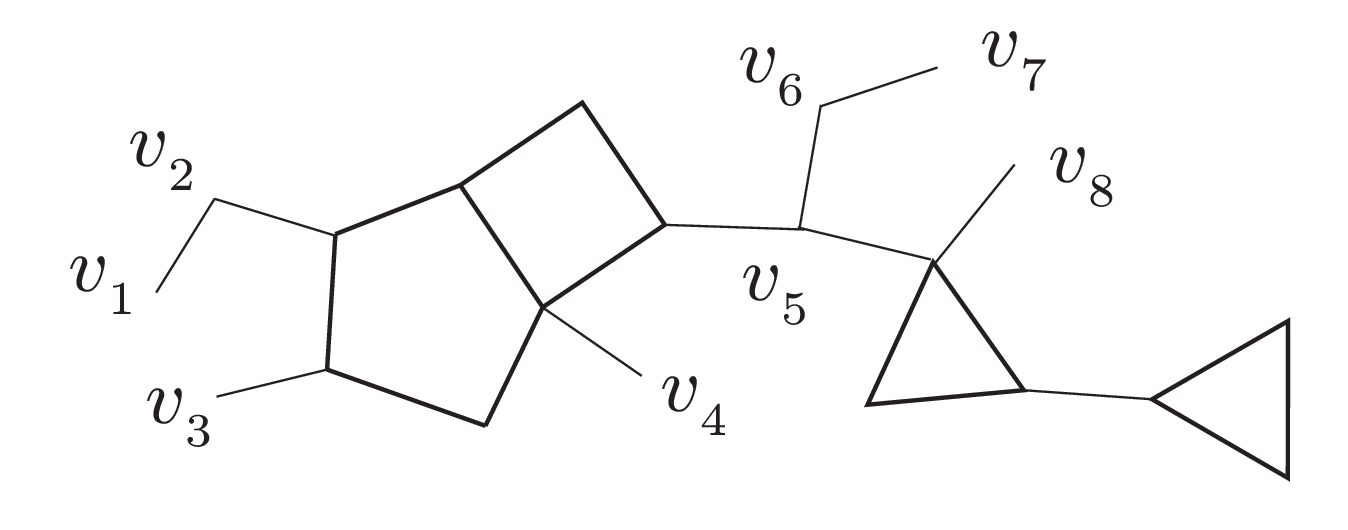} \includegraphics[width=0.42\textwidth]{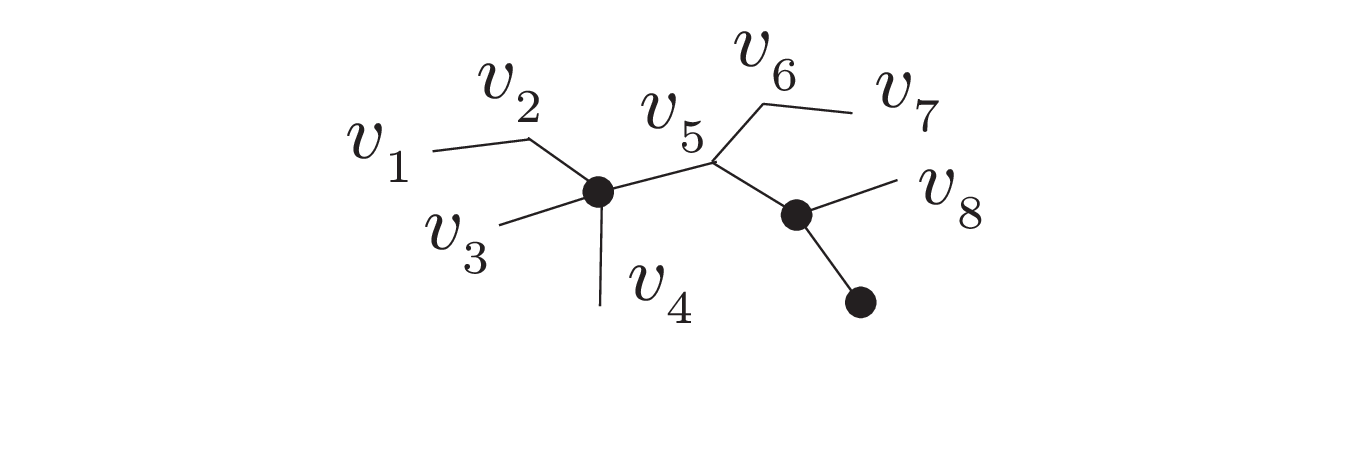} 
\end{center}
\caption{\label{fig:skeleton} The skeleton tree (on the right) after contracting the graph on the left (from Fig. \ref{fig:peqn-1}); the black dots are the composite vertices.}
\end{figure}

Next, we contract $G$ to get a {\em skeleton tree}, $T_G$, by collapsing each TECC into a {\em composite vertex}; other vertices and edges are left intact. For example, the graph from Fig. \ref{fig:peqn-1} have the skeleton tree shown in Fig. \ref{fig:skeleton}. This procedure induces a natural map $f_T$ that takes any subgraph $H$ of $G$ to $f_T(H)$ as a subgraph of $T_G$ (via mapping all vertices belonging to the same TECC of $G$ to a composite vertex of $T_G$ and non-composite vertices of $G$ to non-composite vertices of $T$). Given an instance $(G, S, D)$ of \rpp\, with $p < n - 1$ pebbles, all pebbles on the same TECC are equivalent by Lemma \ref{l:spoke}. This induces a problem instance $(T_G, S', D')$ in which all pebbles (in $S$ and $D$) on the same TECC of $G$ are combined into a {\em composite} pebble (in $S'$ and $D'$). Given two vertices $u$ and $v$ in a graph, $u \leadsto v$ denotes a (shortest) path between $u$ and $v$. Such a path is unique when the graph is a tree. By all vertices on (resp. in) $u \leadsto v$, we mean vertices of $u\leadsto v$ including (resp. excluding) $u$ and $v$. Lemma 6 from \cite{AulMonParPer99} can be extended to \rpp\, as follows. 
\begin{lemma}\label{l:exchange} Let $(G, S, D)$ be an instance of \rpp\, in which $G$ is not a cycle and $N(TECCs)$ $\le p < n - 1$. Let $u, v$, and $w$ be vertices of $G$ such that the path between $u$ and $v$ and the path between $v$ and $w$ are not edge disjoint. Assume $u$ and $v$ are occupied by pebbles and moves exist that take $S$ to a new configuration in which pebble $S^{-1}(u)$ is moved to $v$ and $S^{-1}(v)$ is moved to $w$. Then $S$ can be taken to an configuration $S'$ in which $S$ and $S'$ are the same except pebbles on $u$ and $v$ are exchanged. 
\end{lemma}
\ifdefined\FULLPAPER
\noindent {\sc Proof.} For convenience, let $p_1 := S^{-1}(u)$ and $p_2 := S^{-1}(v)$. Let the overlapping part of $u\leadsto v$ and $v \leadsto w$ be $y \leadsto v$. Let the sequence of moves that take $p_1$ to $v$ and $p_2$ to $w$ be represented as $X = \langle S = S_0, S_1, \ldots, D \rangle$. If it is possible to move $p_1, p_2$ to the same TECC, then clearly the locations of $p_1, p_2$ can be exchanged on the TECC without changing any other pebble's configuration. Reversing earlier moves then exchanges $p_1, p_2$ on $u$ and $v$. For the rest of this proof, we assume that $p_1, p_2$ can never occupy vertices from the same TECC. Note that this implies hat $p_1$, $p_2$ can never occupy vertices of the same TECC in different configurations originated from $S$; in particular, no vertex on $y \leadsto v$ can be on a TECC. To see this, if $p_1, p_2$ both reach a TECC $H$ in some (possibly different) configurations in $X$, assume without loss of generality that $p_1$ reaches $H$ first. Since all pebbles on $H$ are equivalent and $H$ contains at least three vertices, $p_1$ can always stay on $H$: Suppose $X$ at some point wants to move $p_1$ outside of $H$. If $p_1$ is the only pebble on $H$, $p_1$ does not hinder any other pebbles from moving through $H$ and moving $p_1$ out will only crowd the rest of $G$, making further pebble movements outside $H$ harder. If $p_1$ is not the only pebble on $H$, we may pick any pebble on $H$ to leave $H$ instead of $p_1$. Then $p_2$ will eventually reach $H$ with $p_1$ still on $H$, allowing them to exchange. 

For the case in which $p_1, p_2$ never visits the same TECC of $G$, let $W$ denote the graph formed by the vertices and edges traveled by $p_1, p_2$ as they move along the sequence of configurations in $X$. Let $T_W = f_T(W)$. If $T_W$ contains  composite vertices that are not leaves of $T_W$, let $z$ be such a composite vertex and $H_z$ be the TECC corresponding to $z$ in $G$. Let $G(H_z, v)$ denote the connected component of $G$ containing $v$ after deleting $H_z$ and let $\overline{G}(H_z, v)$ denote rest of the components. By assumption, only one of $p_1$ or $p_2$ may visit $H_z$. Assume it is $p_1$ (the case of $p_2$ is similar), then $p_2$ can only visit vertices of $G(H_z, v)$; in fact the entire path $v \leadsto w$ is within $G(H_z, v)$. Using the same argument from the previous paragraph, $X$ can be modified so that $p_1$ does not visit vertices of $\overline{G}(H_z, v)$, unless $u \in \overline{G}(H_z, v)$. In this case, however, $p_1$ is equivalent to any pebble that is initially on $H_z$; the lemma holds if an only if a pebble initially on $H_z$ in $S$ can move to $v$ and $p_2$ can move to $w$. Via induction, it must be possible for some pebbles $p_1'$, equivalent to $p_1$, and $p_2$ to move from some $u'$ to $v$ and $v$ to some $w'$, respectively, where $y \leadsto v$ is contained within $u' \leadsto v$ and $v \leadsto w'$. Further more, $p_1', p_2$ do not ``pass through'' any TECC of $G$. 

We may then assume that from the beginning, $T_W$ has only composite vertices that are leaves. Denote the branch of $G$ containing $y$ as $T_y$. Since $p_1, p_2$ may still visit some TECCs, let $T_y'$ denote the tree containing $T_y$ as well as the vertices of these TECCs (visited by $p_1$ or $p_2$) that are (distance 1) neighbors of $T_y$. Since the labels of pebbles other than $p_1, p_2$ have no effect on moving $p_1, p_2$, we may assume pebbles other than $p_1, p_2$ are unlabeled (indistinguishable). It can be shown that unlabeled pebbles outside of $T_y'$ never need to move to $T_y'$: If an unlabeled pebble moves from outside $T_y'$ and stays on $T_y'$ it only makes moving $p_1, p_2$ less feasible; if an unlabeled pebble moves from one vertex outside $T_y'$ to another vertex outside $T_y'$ via $T_y$, it does not help the feasibility of moving $p_1, p_2$ on $T_y'$. Thus, unlabeled pebbles may only move away from $T_y'$ and they should never come back. Therefore, we may first take the unlabeled pebbles that will leave $T_y'$ and move them outside $T_y'$ in the beginning. After these steps, the initial problem is reduced to moving $p_1$ from $u$ to $v$ and $p_2$ from $v$ to $w$ on the tree $T_y'$; by Lemma 6 from \cite{AulMonParPer99}, $p_1, p_2$ are equivalent. Note that this implies that if $p_1$ (resp. $p_2$) can visit a TECC, then $p_2$ (resp. $p_1$) can visit that TECC as well; it is not possible that a given TECC can only be visited by one of the pebbles from $p_1, p_2$. ~\qed
\else
We omit the proof of Lemma \ref{l:exchange} here which does not add much to the discussion. 
\fi
Lemma \ref{l:exchange} leads to a generalized version of Theorem 4 from \cite{AulMonParPer99} to \rpp, given below. We omit the proof since it is nearly identical (we need extended versions of Corollary 1 and 2 from \cite{AulMonParPer99}, which can be easily proved in the same way Lemma \ref{l:exchange} is proved). 

\begin{theorem}\label{t:permutation} An \rpp\, instance, $(G, S, D)$, in which $G$ is not a cycle and $N(TECCs) \le p < n - 1$, is feasible if and only if the individual exchanges between pebble $i$ and $S^{-1}(D(i))$, $1 \le i \le p$, can be performed using moves without affecting the configurations of any other pebble. 
\end{theorem}

By Theorem \ref{t:permutation}, if an instance of \rpp, $I = (G, S, D)$, is feasible, then pebbles $i$ and $\sigma_{S,D}(i) = S^{-1}(D(i))$ can be exchanged with no net effect on other pebbles. This enables a feasibility test of \rpp\, problems (and therefore, \pmr\, problems): vertices occupied by pebbles are partitioned into equivalence classes such that two pebbles can be exchanged if and only if the vertices occupied by them belong to the same equivalence class. In fact, we apply the $Mark$ algorithm from \cite{AulMonParPer99} on the skeleton tree $T_G$ without any change at the pseudocode level (see \cite{AulMonParPer99} for the simple algorithm description); the main difference is how to check whether two adjacent pebbles are equivalent (Lemma 8 from \cite{AulMonParPer99}). 

Before stating our version of the lemma, some notations are in order. We work with an arbitrary \rpp\, instance $I = (G, S, D)$ in which $G$ is not a cycle and $N(TECCs) \le p < n - 1$. Let $I' = (T_G, S', D')$ be the induced instance described earlier in which $T_G$ is $G$'s skeleton tree. A {\em fork} vertex of $T_G$ is a vertex of degree at least 3 that is not a composite vertex. $F(u)$ is the set of connected components of $T_G$ after deleting the vertex $u$. $T(u, v)$ is the tree of $F(u)$ containing the vertex $v$; $\overline{T}(u, v)$ is the rest of $F(u)$. For two vertices $u, v \in V(T_G)$, $d(u, v)$ is the length of $u \leadsto v$. In the lemmas that follow, only start configuration $S'$ is operated on; same procedure can be applied to $D$. First we need a version of Corollary 3 from \cite{AulMonParPer99} to account for composite vertices; we omit the essentially same proof but point out that although both fork and composite vertices can help two pebbles switch locations, a composite vertex can do so with one fewer empty vertex. 

\begin{lemma}\label{l:neq} Let $p_1 := S'^{-1}(u)$, $p_2 := S'^{-1}(v)$ for $u, v \in V(T_G)$ such that $u \leadsto v$ contains no other pebbles; all vertices on $u \leadsto v$ are of degree 2. Let $w$ be a composite or fork vertex such that $u$ is in $w \leadsto v$. The tree $T(u, w)$ has no more than $d(w, u)$ (resp. $d(w, u) + 1$) empty vertices when  $w$ is a composite (resp. fork) vertex. Let $w'$ be the closest composite or fork vertex to $v$ such that $v$ is in $w' \leadsto u$ satisfying similar properties as $w$. Then $u$ and $v$ are not equivalent. 
\end{lemma}
\vspace*{-4mm}
\begin{lemma}\label{l:8} Let $p_1 := S'^{-1}(u)$, $p_2 := S'^{-1}(v)$ for some $u, v \in V(T_G)$ such that $u \leadsto v$ contains no other pebbles. Then $p_1, p_2$ are equivalent with respect to $S'$ if and only if at least one of the following conditions holds:
\begin{list}{}{\leftmargin=0em}
\item 1. There exists a fork vertex $w$ in $u\leadsto v$ such that both $T(w, u), T(w, v)$ are not full or at least one other tree of $F(w)$ is not full. 
\item 2. Let $w$ be a composite vertex such that $u$ is in $w \leadsto v$ and no other fork vertex or composite vertex is in $w \leadsto u$. There exists such a $w$ that $T(u, w)$ has $d(w, u) + 1$ empty vertices. 
\item 3. Symmetric to 2 with $u$ and $v$ switched. 
\item 4. Let $w$ be a fork vertex such that $u$ is in $w \leadsto v$ and no other fork vertex or composite vertex is in $w \leadsto u$. There exists such a $w$ that $T(u, w)$ has $d(w, u) + 2$ empty vertices. 
\item 5. Symmetric to 4 with $u$ and $v$ switched. 
\item 6. Vertex $u$ is a fork vertex. Then at least two trees of $F(u)$ has empty vertices or there are at least two empty vertices outside $T(u, v)$. 
\item 7. Symmetric to 6 with $u$ and $v$ switched. 
\item 8. Vertex $u$ is a composite vertex. Then at least one tree of $\overline T(u, v)$ has an empty vertex. 
\item 9. Symmetric to 8 with $u$ and $v$ switched. 
\end{list}
\end{lemma}
\noindent {\sc Proof.} The proof is adopted from that of Lemma 8 from \cite{AulMonParPer99} with some repetitive details omitted. Since the sufficiency of the conditions can be easily checked by constructing plans that exchange $p_1, p_2$, only necessity is shown here via contradiction. Assume that $u$ and $v$ are exchangeable without configuration $S$ satisfying any of the conditions 1-9. First consider the case in which there is no fork vertex in $u \leadsto v$ and $u$ and $v$ are not fork or composite vertices; these assumptions forbids conditions 1 and 6-9. If conditions 2-5 do not hold, the condition from Lemma \ref{l:neq} is true, thus $u$ and $v$ cannot be equivalent. 

For the case in which no fork vertex exists in $u \leadsto v$ but $u$ or $v$ (possibly both) is a fork or composite vertex, the proof from Lemma 8 from \cite{AulMonParPer99} applies with little change to show that $u$ and $v$ are not equivalent unless one of conditions 2-9 holds: If conditions 2-5 do not hold, this means that $p_1, p_2$ must use $u$ or $v$ as a ``hub'' for switching locations; traveling beyond distance 1 from $u \leadsto v$ will not help $u$ and $v$ to switch. On the other hand, if conditions 6-9 do not hold, $u$ or $v$ cannot serve as the hub that enables $u$ and $v$ to switch. Furthermore, if conditions 6-9 do not hold, reconfiguration of pebbles will not make conditions 2-5, previously invalid, become valid. 

This leaves the case in which conditions 2-9 do not hold, which means that $u$ and $v$ cannot switch on $\overline T(u, v)$ nor $\overline T(v, u)$. Since there is no pebble in $u \leadsto v$, the vertices in $u \leadsto v$ cannot be composite vertices. The same proof from Lemma 8 from \cite{AulMonParPer99} then shows that unless condition 1 is met, $u$ and $v$ cannot be equivalent. ~\qed 

With Lemma \ref{l:8}, all criteria needed for the $Mark$ algorithm from \cite{AulMonParPer99}, in particular Observations 1-4, continue to hold on $T_G$ without change. Since $Mark$ is not changed, its running time is linear if deciding whether two adjacent pebbles are equivalent can be performed in (amortized) constant time. For this to hold, for an arbitrary tree $T(u, w)$, we need to know whether $T(u, w)$ has 0, 1, 2 holes and whether the fork or composite vertex of $T(u, w)$ closest to $u$ allows $u$ and another vertex $v$ in $T(u, w)$ to exchange ({\em i.e.}, $T(u, w)$ should have enough empty vertices). These data can be precomputed in $O(|V| + |E|)$ time using two depth firth traversals over the tree $T_G$. At this point, it is not hard to see that this linear decision algorithm easily turns into an algorithm that computes a feasible solution to a \ppr\, instance. Our complexity analysis shows that a feasible solution can be computed in $O(|E|)$ if a high level plan is required (computes a corresponding \rpp\, instance, checks feasibility, and outputs the permutation pairs for exchanges) and $O(n^3)$ if step by step output is required (each exchange can be done in $O(n^2)$ moves produced by a fixed formula). We summarize the main result of this section with the following theorem.
\begin{theorem}The feasibility of \pmr\, problems can be decided in linear time. Moreover, a plan for a feasible instance can be computed in $O(n^3)$ time.\end{theorem}

\vspace*{-6mm}
\section{Conclusion}\label{sec:conclusion}
\vspace*{-4mm}
In this paper, we proposed the problem of {\em pebble motion on graphs with rotations} (\pmr), a graph-based multi-robot path planning problem. Our formulation takes into account natural, synchronous rotations of pebbles along fully occupied cycles of the underlying graph. The inclusion of this important case, in conjunction with previous studies of the problem that only allow pebbles to move to unoccupied vertices, paints a fairly complete picture of graph-based multi-robot path planning problems. In our systematic analysis of \pmr, we show that, even for the fully constrained case in which the number of pebbles equals the number of vertices, deciding the feasibility of a \pmr\, instance can be completed in linear time with respect to the size of the underlying graph. Moreover, computing a full plan for all moving all pebbles requires $O(n^3)$ time. 


\bibliographystyle{IEEETranN}
\bibliography{publications,references,references2}


\end{document}